\documentclass[12pt]{article}
\usepackage{graphicx,epsfig,amsfonts,amssymb}
\begin{document}

\newcommand{\be}{\begin{equation}}
\newcommand{\ee}{\end{equation}}
\newcommand{\bea}{\begin{eqnarray}}
\newcommand{\eea}{\end{eqnarray}}
\newcommand{\beas}{\begin{eqnarray*}}
\newcommand{\eeas}{\end{eqnarray*}}

\begin{titlepage}
\begin{flushright}
{\small CU-TP-1086} \\
{\small hep-th/0307233}
\end{flushright}

\begin{center}

\vspace{5mm}

{\Large \bf Brane gases in the early universe: \\ thermodynamics and
cosmology}

\vspace{3mm}

Richard Easther${}^a$\footnote{\tt easther@physics.columbia.edu},
Brian R.~Greene${}^{ab}$\footnote{\tt greene@physics.columbia.edu},
Mark G.~Jackson${}^c$\footnote{\tt markj@physics.columbia.edu} \\
and Daniel Kabat${}^c$\footnote{\tt kabat@physics.columbia.edu}
\setcounter{footnote}{0}

\vspace{2mm}

${}^a${\small \sl Institute for Strings, Cosmology and Astroparticle
Physics} \\
{\small \sl Columbia University, New York NY 10027}
\vspace{1mm}

${}^b${\small \sl Department of Mathematics} \\
{\small \sl Columbia University, New York, NY 10027}
\vspace{1mm}

${}^c${\small \sl Department of Physics} \\
{\small \sl Columbia University, New York, NY 10027}

\end{center}

\vskip 0.3 cm

\noindent
We consider the thermodynamic and cosmological properties of brane
gases in the early universe. Working in the low energy limit of
M-theory we assume the universe is a homogeneous but anisotropic
10-torus containing wrapped 2-branes and a supergravity gas.  We
describe the thermodynamics of this system and estimate a Hagedorn
temperature associated with excitations on the branes.  We investigate
the cross-section for production of branes from the thermal bath and
derive Boltzmann equations governing the number of wrapped branes.  A
brane gas may lead to decompactification of three spatial dimensions.
To investigate this possibility we adopt initial conditions in which
we fix the volume of the torus but otherwise assume all states are
equally likely.  We solve the Einstein-Boltzmann equations
numerically, to determine the number of dimensions with no wrapped
branes at late times; these unwrapped dimensions are expected to
decompactify.  Finally we consider holographic bounds on the initial
volume, and find that for allowed initial volumes all branes typically
annihilate before freeze-out can occur.

\end{titlepage}

\section{Introduction}

One of the few firm predictions of string or M-theory is the existence
of extra spatial dimensions.  The conventional scenario is that these
dimensions are unobservably small today.  Understanding how the
universe got into such an asymmetric state is necessarily a problem
for cosmology.  One intriguing possibility is that extended objects
play a vital role in explaining the asymmetry.  This is the idea
behind {\em brane gas cosmology\/} \cite{Alexander:2000xv,
Brandenberger:2001kj,Easson:2001fy,Watson:2002nx,Boehm:2002bm,
Alexander:2002gj}.

Brane gas cosmology rests on the assumption that in string or M-theory
branes will be present as one component of a heat bath that fills the
early universe.  In a universe with compact directions that can be
wrapped by branes, the dynamics of the wrapped branes may play a
significant role in the overall evolution of the universe. In
particular one can explore analogues of the Brandenberger-Vafa
scenario, which proposes that the three large spatial dimensions we
see at present arose from a thermal fluctuation in a primordial gas of
winding strings \cite{Brandenberger:1988aj,Tseytlin:1991xk,
Cleaver:1994bw,Sakellariadou:1995vk,Easther:2002mi}.

In a recent paper we considered the late time behavior of a brane gas
model, arising from M-theory compactified on $T^{10}$, in which the
universe contains a supergravity gas and 2-branes wrapped on the
various cycles of the torus \cite{Easther:2002qk}. We could safely
ignore 5-branes, which are also part of the M-theory spectrum, since
they will quickly intersect and annihilate in ten spatial
dimensions. The key conclusion of \cite{Easther:2002qk} was that the
directions which were not wrapped by 2-branes expanded faster than
those that were, and that the overall expansion rate of the wrapped
and unwrapped subspaces depended only on their dimensionality.  In the
present paper we turn our attention to the early time behavior of this
model, and include two crucial ingredients which are unimportant at
late times: the fluctuations on the branes themselves, and the
annihilation and creation of branes out of the thermal bath provided
by the supergravity gas.

This paper is laid out as follows. In the following section we derive
the analogs of the Friedmann equations for this cosmology.  Section 3
examines the statistical mechanics of a universe which contains
2-branes and radiation (the supergravity gas), and presents a simple
and, to our knowledge, novel derivation of a limiting (Hagedorn)
temperature for the 2-brane gas.  In section 4 we investigate the
cosmology of the ``Hagedorn phase,'' showing that in this phase the
negative pressure of the brane tension cancels the positive pressure
of the brane fluctuations, so the universe expands almost as if it
were filled with pressureless dust.  In section 5 we investigate the
cross-section for brane-antibrane annihilation and write down
Boltzmann equations governing the number density of the different
brane wrapping modes. We show that the effective interaction rate
drops to zero at a finite time, producing a {\em freeze-out\/}
analogous to that leading to a relic abundance of dark matter species
in conventional cosmology, so that some directions can remain wrapped
at late times. In Section 6 we numerically evolve the combined
Einstein-Boltzmann equations for a variety of different initial
conditions, and show that the number of directions which are not
wrapped by branes depends on the initial volume of the universe.
Section 7 describes holographic constraints on the initial conditions
for the universe, and we conclude in Section 8.  Throughout this paper
we scale the M-theory Planck length to unity, so that Newton's
constant is given by $16 \pi G = (2\pi)^8$ and the 2-brane tension is
$T_2 = 1/(2\pi)^2$.

\section{Gravitational dynamics}
\label{gravitydynamics}

We consider a universe whose spatial topology is a $d$-dimensional
torus $T^d$.  The case of interest for M-theory is $d=10$, but we
frequently write $d$ as a parameter, to clarify the origin of the
numerical constants appearing in our equations. The universe is thus
spatially flat but has finite volume, a fact which will be of crucial
importance in what follows. We use the metric
\be
\label{metric}
ds^2 = - dt^2 + \sum_{i=1}^d e^{2 \lambda_i(t)} dx_i^2 \qquad 0 \leq
x_i \leq 1
\ee
and add matter in two forms: M2-branes and a supergravity gas.  The
stress tensor for the massless supergravity gas is
\be
{T^\mu}_\nu = {\rm diag} (-\rho_{S},p_{S}, \ldots, p_{S}) 
\ee
where $\rho_S = c_S T^{11}$ is the energy density of the supergravity
gas.  The coefficient $c_S$ is computed in appendix A, and the
equation of state fixes $p_{S} = \frac{1}{d} \rho_{S}$.

The M2-brane gas consists of 2-dimensional membranes, each of which is
wrapped around a 2-cycle inside the $T^{10}$.  These wrapped branes
are thus topologically stable.  There are 45 ($=10\times 9/2$)
independent ways to wrap the 2-branes, so we effectively have 45
distinct species of branes in our model.  The universe is spatially
compact, so to satisfy Gauss' law we must have equal numbers of branes
and antibranes on each 2-cycle.

Note that we are ignoring the possibility of having ``diagonally
wound'' branes, which wrap on non-trivial linear combinations of the
basic $(ij)$ homology cycles.  This is required by our metric ansatz
(\ref{metric}) which describes a rectangular torus.  Under time
evolution such a torus is compatible with having branes wound in the
$(ij)$ directions, but not with having diagonally wound branes, whose
tension would cause the torus to tilt.

The brane energy density gets contributions from two sources: the brane
tension and the energy in transverse brane fluctuations.  We ignored
the latter contribution in Ref.~\cite{Easther:2002qk} since it is
negligible at late times, but we must include it here.  We describe
the wrapped branes using the leading long-wavelength approximation to
the Nambu-Goto action, and thus model the transverse fluctuations as a
non-interacting gas of massless particles living on the brane.  The
entire brane can also move in the transverse directions, but we assume
that this motion is non-relativistic.  The brane is thus effectively
at rest, so its kinetic energy is negligible compared to its rest mass
and can be ignored.

For a brane at rest wrapped once around the $(12)$ cycle and smeared
over the eight transverse dimensions, the contribution to the stress
tensor from the brane tension is
\be {T^\mu}_\nu = -\frac{T_2}{ {\rm vol}_\perp} {\rm diag} 
(1,1,1,0,\ldots,0), 
\ee
where $T_2$ is the brane tension and ${\rm vol}_\perp =
\exp{\sum_{i=3}^{10} \lambda_i}$ denotes the volume perpendicular to
the brane.  Similarly, for the worldvolume gas on the brane
\be 
{T^\mu}_\nu = \frac{1}{{\rm vol}_\perp} {\rm diag}
(-\rho_X,p_X,p_X,0,\ldots,0),
\ee
where $\rho_X = c_X T^3$ is the energy density in the fluctuation gas.
The coefficient $c_X$ is computed in appendix A.  The equation of
state for a gas in two spatial dimensions fixes $p_X = \frac{1}{2}
\rho_X$.

Combining these expressions we can write down the energy density for
this universe:
\be
\label{rhofull}
\rho = c_S T^{11} + {1 \over V} \sum_{i \not= j} N_{ij} (T_2 + c_X T^3)
e^{\lambda_i + \lambda_j}\,.
\ee
Here $N_{ij}$ for $i>j$ is the number of branes wrapped on the $(ij)$
cycle, $N_{ji} = N_{ij}$ is the number of antibranes, and $V = e^{\sum
\lambda_i}$ is the total volume of the torus.  Likewise the pressure
on the $i^{th}$ dimension is
\be
\label{pfull}
p_i = {1 \over d} c_S T^{11} + {1 \over V} \sum_{k \not= i}
\left(N_{ik} + N_{ki}\right) \left(-T_2 + {1 \over 2} c_X T^3\right)e^{\lambda_i + \lambda_j}\,.
\ee
The relevant Einstein equations are derived in appendix B.  They take
the form of a Hamiltonian constraint (the analog of the Friedmann
equation)
\be
\label{Hconstraint}
{1 \over 16 \pi G} \sum_{i \not= j} \dot{\lambda}_i \dot{\lambda}_j = \rho
\ee
along with a set of dynamical equations of motion
\be
\label{lambdaddot}
\ddot{\lambda}_i + (\sum_k \dot{\lambda}_k) \dot{\lambda}_i = 8 \pi G \left({1 \over d-1} \rho + 
p_i - {1 \over d-1} \sum_k p_k\right)\,.
\ee
%

\section{Brane gas thermodynamics}

In this section we work out the statistical distribution of scale
parameters $\lambda_i$, velocities ${\dot \lambda}_i$ and wrapping
numbers $N_{ij}$ when the system is in thermal equilibrium.

\subsection{Empty universes}

We start with the trivial case of an empty universe, with no
supergravity gas and no branes, and begin by setting up the canonical
formalism.  For the metric $(\ref{metric})$, the curvature scalar is
\be
R = - 2 \sum_i \ddot{\lambda}_i - 2 \sum_i \dot{\lambda}_i^2
- \sum_{i \not= j} \dot{\lambda}_i \dot{\lambda}_j
\ee
and the Einstein-Hilbert action is
\bea
\nonumber
S & = & - {1 \over 16 \pi G} \int d^{11}x \sqrt{-g} \, R \\
\label{action}
& = & - {1 \over 16 \pi G} \int dt \, V
\sum_{i \not= j} \dot{\lambda}_i \dot{\lambda}_j
\eea
where $V = e^{\sum_i \lambda_i}$ and we have integrated by
parts in the second line.  The canonical momenta $\pi_i$ are given by
\be
\label{canmom}
\pi_i = {\partial {\cal L} \over \partial \dot{\lambda}_i}
= - {V \over 8 \pi G} \sum_{j \not= i} \dot{\lambda}_j
\ee
and the Hamiltonian is
\be
\label{Hamiltonian}
H_{\rm gravity} = \sum_i \pi_i \dot{\lambda}_i - {\cal L} 
= - {V \over 16 \pi G} \sum_{i \not= j}
\dot{\lambda}_i \dot{\lambda}_j \,.
\ee
The equations of motion one obtains from the action (\ref{action}) or
the Hamiltonian (\ref{Hamiltonian}) do not completely reproduce the
Einstein equations, since our metric (\ref{metric}) fixes a choice of
gauge $g_{tt} = -1$.  Varying $g_{tt}$ gives the time-time component
of the Einstein equations, which is a constraint that must be imposed
on the initial conditions.  One can check that this constraint is
equivalent to requiring that the Hamiltonian vanishes, $H_{\rm
gravity} = 0$.  This condition is, of course, expected in a spatially
compact universe.  With this constraint the equations of motion that
follow from the Hamiltonian (\ref{Hamiltonian}) are equivalent to the
usual Einstein equations.  This constraint has been interpreted to
mean that the ``wave function of the universe" should satisfy $H_{\rm
gravity} \Psi = 0$ \cite{Dewitt,wheelerdewitt, Hartle:1983ai}.

Now consider the equilibrium distribution of states for an empty
universe.  Quantizing the system semiclassically, and assuming that
all zero-energy states are equally likely, the volume of phase space
available to the system is
\be
\Gamma = \int {d^d \pi \, d^d\lambda \over (2 \pi)^d} \,
\delta(H_{\rm gravity})\,.
\ee
This is nothing but the microcanonical ensemble of classical statistical
mechanics.  It is more transparently written in terms of the radii
$R_i = {1 \over 2 \pi} e^{\lambda_i}$ as
\be
\Gamma = {1 \over 4 \pi G} \int d^d \pi \, d^d R \,\,
\delta\Bigl(\sum_i \pi_i^2 - {1 \over d-1} (\sum_i \pi_i)^2\Bigr)\,.
\ee
Note that the radii are uniformly distributed from zero to infinity.
That is, in equilibrium the typical universe has very large volume and
is very anisotropic.

\subsection{Adding matter}
\label{AddMatter}

We consider three different matter contributions to the energy of the
system, namely
\be
E_{\rm matter} = E_S + E_T + E_X
\ee
arising from the supergravity gas, brane tension, and excitations on
the branes, respectively (note the slight abuse of the term `matter'
to include the radiation-like supergravity gas).  At temperature $T$
the energy and entropy of the supergravity gas are given by
\bea
\nonumber
E_S & = & c_S V T^{11} \\
\label{SugraEnt}
S_S & = & {11 \over 10} c_S V T^{10}\,.
\eea
The coefficient $c_S$ is worked out in appendix A.  For branes at rest
the energy due to brane tension is
\be
E_T = T_2 \sum_{i \not= j} N_{ij} e^{\lambda_i + \lambda_j}
\ee
while the energy and entropy due to a massless gas of excitations on
the branes are given by
\bea
\nonumber
E_X & = & \sum_{i \not= j} N_{ij} c_X e^{\lambda_i + \lambda_j} T^3 \\
\label{BraneEnt}
S_X & = & \sum_{i \not= j} N_{ij} {3 \over 2} c_X e^{\lambda_i +
\lambda_j} T^2\,.
\eea
The coefficient $c_X$ is worked out in Appendix A.

The above expressions for the energy and entropy of massless particles
are only exact in the thermodynamic limit.  But fortunately there are
enough massless quanta in the early universe for these expressions to
be precise.  The thermodynamics of the branes, on the other hand, is
more subtle, because the universe may contain only a small number of
branes at early times, and moreover the branes that we do have are
divided into 45 different sub-populations, labelled by the directions
$(ij)$ on which they are wrapped.  Thus we need to allow for thermal
fluctuations in the brane wrapping numbers $N_{ij}$.

To do this we study the probability distribution for the combined
matter--gravity system.  The volume of phase space is
\be
\label{distribution}
\Gamma = \int {d^d\pi \, d^d\lambda \over (2 \pi)^d} \sum_{N_{ij}}
\int dE_{\rm matter} e^{S_{\rm matter}} \delta(H_{\rm gravity} +
E_{\rm matter})\,.
\ee
Note that this expression corresponds to the microcanonical ensemble
of statistical mechanics, since the $\delta$-function enforces the
constraint that the total energy in the universe must be zero.  Given
that $E_{\rm matter} > 0$, we must have $H_{\rm gravity} < 0$, which
is possible since the gravitational Hamiltonian (\ref{Hamiltonian}) is
unbounded below.  To avoid a possible confusion, note that in
(\ref{distribution}) we are regarding $S_{\rm matter}$ as a function
of $E_{\rm matter}$, as appropriate when working in the microcanonical
ensemble.  That is, in (\ref{SugraEnt}) and (\ref{BraneEnt}) we regard
$T$ merely as a convenient control parameter, which we will determine
shortly in terms of $E_{\rm matter}$.

Using the $\delta$-function to evaluate the integral in
(\ref{distribution}) gives our final result for the distribution of
radii, velocities and wrapping numbers.  Up to inessential numerical
factors
\be
\label{prob}
\Gamma = \int d^d\lambda \, d^d\dot{\lambda} \, \sum_{N_{ij}} \, V^d e^{S_{\rm matter}}\,.
\ee
In this expression the matter entropy is
\be
S_{\rm matter} = {11 \over 10} c_S V T^{10}
+ \sum_{i \not= j} N_{ij} {3 \over 2} c_X e^{\lambda_i + \lambda_j} T^2
\ee
where temperature is fixed by the Hamiltonian constraint
\[
H_{\rm gravity} + E_{\rm matter} = 0\,.
\]
Written out explicitly this constraint reads
\be
-{V \over 16 \pi G} \sum_{i \not= j} \dot{\lambda}_i \dot{\lambda}_j + c_S V T^{11}
+ \sum_{i \not= j} N_{ij} \left(T_2 + c_X T^3 \right) e^{\lambda_i + \lambda_j} = 0\,.
\ee
Incidentally, it is easy to show that the volume of the universe
increases monotonically, by rewriting the Hamiltonian constraint as
\be
\label{monotonic}
\left({\dot{V} \over V}\right)^2 = \sum_i (\dot{\lambda}_i)^2
+ {16 \pi G E_{\rm matter} \over V} \geq 0\,.
\ee
%

\subsection{Maximum entropy configurations}

The distribution (\ref{prob}) is dominated by configurations which
maximize $S_{\rm matter}$.  We now turn to the problem of identifying
these equilibrium configurations.  We will hold the radii of the torus
fixed, so the quantities we can vary are the wrapping matrix $N_{ij}$
and the temperature $T$.  Introducing a Lagrange multiplier $\mu$ to
enforce the Hamiltonian constraint, we wish to extremize
\be
\label{varprob}
S_{\rm matter} - \mu \left(E_{\rm matter} - \frac{V}{16 \pi G}
\sum _{i \neq j} {\dot \lambda}_i {\dot \lambda}_j \right)
\ee
with respect to $T$, $N_{ij}$ and $\mu$.  Extremizing with respect to
$T$ yields
\be
\mu = \left. \frac{\partial S_{\rm matter}}{\partial E_{\rm matter}} \right|_{\lambda_i}
\equiv \frac{1}{T} 
\ee
so that (\ref{varprob}) is proportional to the free energy.
Extremizing with respect to $N_{ij}$ yields an equation that fixes the
temperature of the universe:
\bea
&& c_X \left( \frac{3}{2} T^2 - \mu T^3 \right) - \mu T_2 = 0
\nonumber \\
&& \Rightarrow T = T_H \equiv \left( \frac{2 T_2}{c_X} \right)^{1/3}\,.
\eea  
We will refer to $T_H$ as the M-theory Hagedorn temperature, for
reasons we discuss in more detail in the next section.  Finally
extremizing with respect to $\mu$ enforces the Hamiltonian constraint
(\ref{Hconstraint}), which fixes the equilibrium total area in
membranes to be
\be
\label{Nequi}
{\cal N}_{\rm eq} \equiv \sum_{i \not= j} N_{ij} e^{\lambda_i + \lambda_j} =
{V \over 3 T_2} \left({1 \over 16 \pi G} \sum_{i \not= j} \dot{\lambda}_i
\dot{\lambda}_j - c_S T_H^{11}\right)\,.
\ee
The velocities $\dot{\lambda}_i$ decrease as the universe expands.  At
some point the equilibrium area in membranes goes to zero.  Beyond
this point the right hand side of (\ref{Nequi}) becomes negative,
which simply means that no branes are present in
equilibrium.\footnote{Of course branes which have dropped out of
equilibrium may be present at arbitrarily late times.}  In this regime
only the supergravity gas remains in thermal equilibrium, with a
temperature that monotonically drops below $T_H$.

To move from the equilibrium total area (\ref{Nequi}) to the
equilibrium wrapping matrix itself we assume that, on average, the
membrane area gets equally distributed among all the 2-cycles.  Thus
the equilibrium number of branes wrapped on the $(ij)$ cycle is
\be \label{Nijequi}
\{N_{\rm eq}\}_{ij} = {1 \over d (d-1)} {\cal N}_{\rm eq}
e^{-\lambda_i - \lambda_j}\,.
\ee
Note that in equilibrium large dimensions are wrapped by fewer branes
than small dimensions.

\section{Limiting temperatures in M-theory}
\label{HagSect}

We denoted the critical temperature found in the previous section by
$T_H = (2 T_2 / c_X)^{1/3}$.  Using $T_2 = 1/(2\pi)^2$ and the value
of $c_X$ from appendix A, this temperature has the numerical value (we
usually set $M_{11} = 1$)
\begin{eqnarray}
\label{Mestimate}
T_{H} &=& \left( 28 \pi \zeta (3) \right)^{-1/3} M_{11} \\
\nonumber
&\approx& 0.211473 \ M_{11} \,.
\end{eqnarray}
This is very close to the M-theory critical temperature found by Russo
\cite{Russo:2001vh} after a much less heuristic calculation. Our
result differs from Russo's by a factor of $2^{1/3} \approx 1.2599$.
We discuss this discrepancy in more detail in appendix C.  In any case
we interpret $T_H$ as the \emph{M2-brane limiting temperature}, in the
same sense that the Hagedorn temperature is the string limiting
temperature.

The physics behind this Hagedorn behavior is simply that branes can be
created from the thermal bath.  Suppose we start at low temperature
and increase the matter energy density.  The temperature of the
universe will rise to $T_{H}$ then stay there.\footnote{We are
implicitly assuming that the specific heat does not diverge too
rapidly as $T \rightarrow T_H$.  In string theory this assumption was
studied in \cite{Brandenberger:1988aj}.}  As the matter energy density
increases further, the extra energy will be pumped into the creation
of M2-branes.  Thus the brane number density will increase while the
energy density of the supergravity gas stays fixed.  Conversely, if
the universe starts out in this Hagedorn phase, the matter energy
density decreases as the universe expands.  A point will be reached at
which the equilibrium $N_{ij}$ drop to zero and the configuration
consists entirely of supergravity gas.  Beyond this point $\rho \sim
\rho_S \sim T^{11}$, so for the energy density to drop further the
temperature must drop.

To be complete, we mention \emph{another} limiting temperature,
associated with the presence of M5-branes.  These are not important
for the late-time dynamics of \cite{Easther:2002qk} because they
intersect and annihilate quickly in the 10 spatial dimensions of
M-theory, but their presence at early times implies a limiting
temperature ($T_5 = M_{11}^6/(2\pi)^5$):
\begin{eqnarray}
T_{M5} &=& \left( \frac{5 T_5}{c'_X} \right)^{1/6} \\
\nonumber
&=& \left( \frac{15}{16 \pi^8} \right)^{1/6} M_{11}\\
\nonumber
&\approx& 0.215012 \ M_{11}
\end{eqnarray}
Note that $ T_{H} \approx T_{M5}$.  We do not believe that there are
two limiting temperatures, one for M2-branes and one for M5-branes,
and it is just coincidence that they happen to be nearly identical.
Rather we conjecture that these are indications of a \emph{single}
M-theory limiting temperature $T_M \approx 0.2 M_{11}$.  We believe
that a full understanding of M-theory (including higher-order
corrections to the supergravity action we are studying here) will
produce the corrections necessary for the two temperatures to
coincide.  Indeed, since both temperatures were computed only at
lowest order, it is remarkable they agree this well.  We leave this as
a significant open problem.

Finally, it is interesting to compare our M-theory results to the
limiting temperature expected in string theory.  In general some of
the M2-branes could be wrapped around directions whose scale factors
$e^{\lambda_i}$ and $e^{\lambda_j}$ differ significantly from one
another.  When both scale factors are large we expect
(\ref{Mestimate}) to be a reasonable estimate for the limiting
temperature.  But when one scale factor becomes small the membrane can
be modelled as a string, with the small dimension playing the role of
the dilaton.  For type II strings the limiting Hagedorn temperature is
(expressed in M-theory units)
\be T_H = \left( \pi \sqrt{8 \alpha'} \right) ^{-1} = \frac{1}{\pi 
\sqrt{8}} \sqrt{R_{10}} M^{3/2}_{11} 
\ee
where $R_{10}$ is the radius of the small dimension.  The true
limiting temperature should interpolate between these two extremes; it
would be interesting to study this in more detail.

\subsection{Thermodynamics and cosmology in the Hagedorn phase}

We now consider the evolution of a universe in the Hagedorn phase.
For simplicity we specialize to the case of an isotropic torus (all
$\lambda_i$ equal).  The ``brane area density'' is
\be
n = {\hbox{\rm total area in wrapped branes} \over \hbox{\rm total
volume of the universe}} = \frac{{\cal N}_{\rm eq}}{V}
\ee
where ${\cal N}_{\rm eq}$ is given by equation~(\ref{Nequi}).
This must be positive, which requires
\be
\label{MinHagH}
\dot{\lambda} > \left({16 \pi G c_S T_H^{11} \over d(d-1)}\right)^{1/2} \approx \,\, 0.502 \quad \hbox{\rm in Planck units.}
\ee
Thermodynamics in the Hagedorn phase is straightforward.  As always,
the energy density is fixed by the Hamiltonian constraint,
\be
\label{HamCon}
\rho = c_S T^{11} + {1 \over V} \sum_{i \not= j} N_{ij} (T_2 + c_X T^3) e^{\lambda_i + \lambda_j}
= {d(d-1) \over 16 \pi G} \dot{\lambda}^2
\ee
In general the pressure is given by (\ref{pfull}). At the Hagedorn
temperature the positive pressure due to excitations on the branes
exactly cancels the negative pressure due to brane
tension.\footnote{This can be understood by noting that for an
extensive thermodynamic system the pressure is (minus) the free energy
density.  The latter quantity vanishes for branes at the Hagedorn
temperature.}  Thus the pressure
\be
p = {1 \over d} c_S T_H^{11}
\ee
is isotropic and comes only from the supergravity gas.  Finally the
entropy density is given by
\be
\label{sHag}
s = {1 \over T_H} \left({1 \over d} c_S T_H^{11} + {d(d-1) \over 16
\pi G} \dot{\lambda}^2\right)\,.
\ee
To determine the evolution of the scale factor we proceed in the usual
way.  For an isotropic universe energy conservation requires
\be
d\left(\rho e^{d\lambda}\right) = - p d\left(e^{d\lambda}\right)\,.
\ee
The pressure is constant, so this implies
\be
\rho + p = {{\rm const.} \over e^{d\lambda}}\,.
\ee
Plugging this result into the Hamiltonian constraint (\ref{HamCon})
gives a differential equation for the scale factor.  The general
solution is
\be
e^{\lambda(t)} = {\rm const.} \, \sin^{2/d}\left({t  \over 2}
\sqrt{{16 \pi G d p \over d-1}}\right)
\ee
where we have fixed initial conditions $\lambda \rightarrow -\infty$
as $t \rightarrow 0$. 

This result looks a little odd, but it is only valid when the universe
is in the Hagedorn phase ($T = T_H$), so it cannot be interpreted as
an oscillatory universe. Having found the exact solution, it's
actually an excellent approximation to neglect the pressure.  Recall
that the brane gas does not contribute to the pressure, so in the
Hagedorn phase we have an inequality
\be
p = p_{\rm SUGRA} = {1 \over d} \rho_{\rm SUGRA} \leq {1 \over d} \rho
\ee
In this approximation the universe is filled with pressureless dust,
and the scale factor has the usual matter-dominated form
$e^{\lambda(t)} = {\rm const.} \, t^{2/d}$.

\section{Brane annihilation}

We now look at interactions between the branes and the supergravity
gas, which communicate via the reaction
\be M2{\rm -brane} + \overline{M2}{\rm-brane} \leftrightarrow \ {\rm 
SUGRA}\ {\rm particles}\,.
\ee
For thermodynamic equilibrium the interaction rate must be
sufficiently high, which means the branes must be able to meet in the
transverse dimensions.  Hence interactions will be suppressed if the
transverse dimensions are big, as discussed in
\cite{Brandenberger:1988aj,Tseytlin:1991xk}.  When the interaction
ceases, the branes are ``frozen in," and will remain wound for the
remainder of the cosmological evolution. This process is exactly
analogous to the freeze-out of dark matter in standard cosmology.

To describe this process quantitatively we need the cross-section for
brane-antibrane annihilation.  It is not clear how to calculate this
from first principles in M-theory.  For inspiration we turn to an
analogous process in string theory, namely the annihilation of two
fundamental strings wound on a torus with opposite orientations.  The
basic process was studied by Polchinski \cite{Polchinski:cn}; for more
details on the following calculation the reader should consult his
paper.  For two strings moving in the $x^1$ direction and wrapped with
opposite orientations on $x^2$ the center-of-mass momentum and winding
vectors are
\beas
& & p_1^\mu = (E, E v,0)\,\,\,\,\, \qquad \ell_1^\mu = (0,0, L_2) \\
& & p_2^\mu = (E,-E v,0) \qquad \ell_2^\mu = (0,0,-L_2)
\eeas
where $L_i$ denotes the size of the torus in the $x^i$ direction.
Thus
\be
s_R = -(p_{1R} + p_{2R})^2 = {L_2^2 \over 4 \pi^2 (1 - v^2)}
\ee
where $p_R^\mu = p^\mu + {1 \over 4 \pi} \ell^\mu$.  Following
Polchinski \cite{Polchinski:cn}, the annihilation probability during a
collision is given by the optical theorem
\be
{\rm prob.} = {1 \over v} {\rm Im} T_{ii} =
{1 \over 4 E^2 v} \left({\kappa_{2} \over 2 \pi}\right)^4
{16 \pi^3 \over \kappa^2_{2}} {\rm Im} I(s_R,t_R=0)
\ee
where $\kappa^2_{2}$ is the 1+1 dimensional gravitational coupling
and the imaginary part of the Shapiro-Virasoro amplitude is ${\rm Im}
I(s_R,t_R=0) = 2 \pi^2 s_R^2$.  The strings collide repeatedly, since
$x^1$ direction is periodic, so it is more convenient to work in terms
of the annihilation probability per unit time
\be
\label{stringrate}
{{\rm prob.}\over{\rm time}} = {2 T_1 \kappa^2_{10} \over L_1 \cdots L_9}
L_2^2 f(v) \qquad f(v) = {2 \over 1 - v^2}\,.
\ee
We inserted a factor of $4 \pi T_1$ on dimensional grounds, where
$T_1$ is the fundamental string tension, and expressed the result in
terms of the 9+1 dimensional gravitational coupling $\kappa^2_{10} =
\kappa^2_{2} L_2 \cdots L_9$.  The key qualitative features are that
the annihilation rate is proportional to the gravitational coupling,
inversely proportional to the volume of the torus, and proportional to
the square of the length of the wound strings.

We assume that the annihilation rate for two oppositely-oriented
membranes has similar qualitative features.  We write it as
\be
\label{2memrate}
{{\rm prob.}\over{\rm time}} = {2 T_2^{4/3} \kappa^2_{11} \over V} A^2 f(v)
\ee
where $A$ is the area of the wrapped membranes and $V$ is the volume
of the torus.  The peculiar fractional power of the membrane tension
is required on dimensional grounds.  The string result suggests that
$f(v) = 2/(1-v^2)$; since we are interested in slowly moving membranes
we will take $f(v) \approx {\rm const.} \approx 2$.

It is straightforward to promote the two-membrane annihilation rate
(\ref{2memrate}) to a Boltzmann equation governing the evolution of
the brane wrapping matrix $N_{ij}$.
\be
\label{boltzmann}
{d \over dt} N_{ij} = - {16 \pi G \, T_2^{4/3} \, f \,
e^{2(\lambda_i + \lambda_j)} \over V}
\left((N_{ij})^2 - (N_{ij}^{\rm eq}){}^2\right)\,.
\ee
Here $N_{ij}^{\rm eq}$ is the equilibrium wrapping matrix
(\ref{Nijequi}), and we have used $2 \kappa_{11}^2 = 16 \pi G$.  Branes
wrapped on the $(ij)$ directions will freeze out when their
annihilation rate
\be
\label{Gammaij}
\Gamma_{ij} = \frac{16 \pi G \, T_2^{4/3} \, f \, e^{2(\lambda_i + \lambda_j)} 
N_{ij}}{V}
\ee
is small compared to the Hubble rate, $\Gamma_{ij} \ll H$.  Note that
we take $\Gamma_{ij}$ to be proportional to $N_{ij}$, not $N_{ij}^{\rm
eq}$, so that we get a sensible annihilation rate even when
$N_{ij}^{\rm eq} = 0$.  In practice we say that freeze-out occurs when
the largest $\Gamma_{ij} < 0.01 H$.

We conclude with a few comments on these results.  First, note that
both the string and membrane annihilation rates (\ref{stringrate}),
(\ref{2memrate}) are compatible with the dimension-counting arguments
of Brandenberger and Vafa \cite{Brandenberger:1988aj}.  For example,
if three dimensions of the torus become large then strings wrapped on
the large dimensions will still be able to annihilate: due to the
factor of $L_2^2$ upstairs in (\ref{stringrate}), the wound strings
effectively behave like point particles moving in one large spatial
dimension.  Likewise, if five dimensions become large membranes
wrapped on the large dimensions will behave like point particles
moving in one big dimension, and thus will still be able to
annihilate.

Also note that we have ignored diagonally-wound membranes.  Membranes
wrapped on the $(ij)$ and $(kl)$ cycles could interact, and indeed
could lower their energy, by merging to form a single membrane wrapped
on the linear combination $(ij) \oplus (kl)$.  Such diagonally-wound
membranes are not compatible with our metric ansatz, for reasons
discussed in section \ref{gravitydynamics}.  Moreover, reactions such
as $(ij) + (kl) \rightarrow (ij) \oplus (kl)$ do not get rid of any
conserved winding numbers, unlike the annihilation to supergravity
particles which we considered above.  So we do not expect that
including diagonally wound membranes would qualitatively affect the
nature of our results.

\section{Numerical simulation}
\label{NumericSect}

We are now in a position to solve the combined Einstein-Boltzmann
equations (\ref{lambdaddot}) and (\ref{boltzmann}).  We have
implemented the equations in a {\sc Fortran} code, allowing us to
consider the evolution of $\lambda_i$ and $N_{ij}$ for many different
sets of initial conditions.

At the level of supergravity, we might expect initial conditions to be
drawn at random from a probability distribution corresponding to the
phase space volume derived in section \ref{AddMatter}.
\be
\label{prob2}
\Gamma = \int d^d\lambda \, d^d\dot{\lambda} \, \sum_{N_{ij}} \, V^d
e^{S_{\rm matter}}
\ee
This amounts to assuming that no state in the early universe is {\em a
priori} special.  In practice, however, we must place some
restrictions on the states we consider.  The first restriction is that
we must fix the initial volume of the universe.  That is, we sample
from the distribution (\ref{prob2}) on a hypersurface with a fixed
value of $\log V = \sum_i \lambda_i$.  In principle this might not
seem like a serious restriction.  The volume increases monotonically
with time, as shown in (\ref{monotonic}), so this is equivalent to
choosing an initial instant of time.  However in practice our results
will depend rather sensitively on the instant of time when we first
assume that semiclassical M-theory is valid and that the universe is
in thermal equilibrium.

A second restriction arises because our action is only a low energy
approximation to M-theory, so it only makes sense to begin studying
the evolution at a moment when this approximation is reasonable.  The
low-energy approximation is valid when all length scales in the
problem are larger than the Planck scale.  We actually have two length
scales associated with each direction -- the physical size
$e^{\lambda_i}$, and the ``Hubble length'' $1/\dot{\lambda}_i$.

The choice of a minimum physical size is not particularly crucial; for
simplicity we will assume that we can trust our action when all
$\lambda_i > 0$.  The choice of a minimum Hubble length is somewhat
more subtle.  In sampling from the distribution (\ref{prob2}) our
results will be dominated by configurations which maximize $S_{\rm
matter}$.  Given the entropy density in the Hagedorn phase
(\ref{sHag}), note that $S_{\rm matter}$ is proportional to the volume
(which we are holding fixed) but is also an increasing function of
$\sum_{i \not= j} \dot{\lambda}_i \dot{\lambda}_j$.  Thus for entropic
reasons our results will be dominated by configurations in which all
$\dot{\lambda}_i$ are equal and as large as possible (equal to the
maximum allowed value).  In a way this is very encouraging, since it
means it is natural for the universe to start out in the Hagedorn
phase.  We will study the dependence on the initial velocities below,
and find that the exact choice of cut-off doesn't make a significant
difference, provided the initial $\dot{\lambda}_i$ are large enough
that the universe begins at the Hagedorn temperature.

For given values of $\lambda_i$ and $\dot{\lambda}_i$ we also need to
specify the matter content.  Assuming we begin in the Hagedorn phase,
the behavior discussed in section \ref{HagSect} means the energy
density of the supergravity gas is equal to $c_S T_H^{11}$.  Any
additional contribution to the energy budget of the universe will be
supplied by branes.  The equilibrium total area in branes is given in
(\ref{Nequi}), so the only remaining question is how to distribute
this area across the different wrapping modes. We do this by assuming
a uniform distribution for the wrapping numbers $N_{ij}$, subject only
to the constraint (\ref{Nequi}).

We now look at two different sets of solutions.  In the first set we
start with all $\dot{\lambda}_i = 1$, and vary the initial volume of
the universe.  In the second set we fix the initial volume and vary
the initial velocities.  In both cases we are interested in
determining the number of directions that are unwrapped at late times.
Our prescription is that we round the wrapping numbers $N_{ij}$ to the
nearest integer at freeze-out.  Thus we say the $i$-th direction is
unwrapped if $N_{ij} < 0.5$ for all $j\not=i$ at the time of
freeze-out.

\subsection{Volume dependence}

We begin by studying how the number of unwrapped dimensions at
freeze-out depends on the initial volume.  To do this we select the
$\lambda_i$ at random, subject to the constraint that $\lambda_i > 0$
and that $\log{V} = \sum_i \lambda_i = {\rm constant}$.  We take all
initial velocities $\dot{\lambda}_i$ equal to unity, and distribute
the wrapping numbers as described above.

\begin{figure}
\begin{center}
\begin{tabular}{cc}
\includegraphics[width=2.5in]{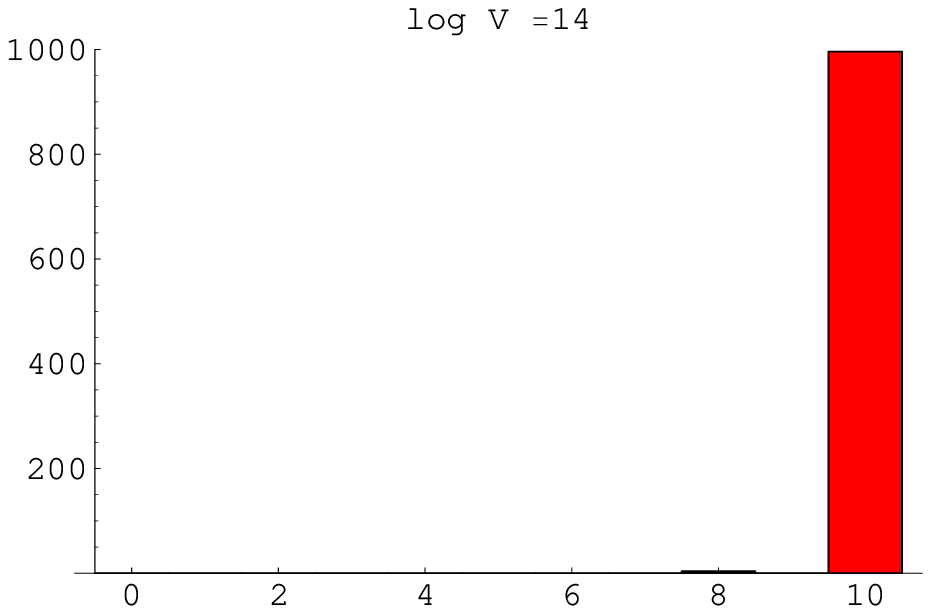} & 
\includegraphics[width=2.5in]{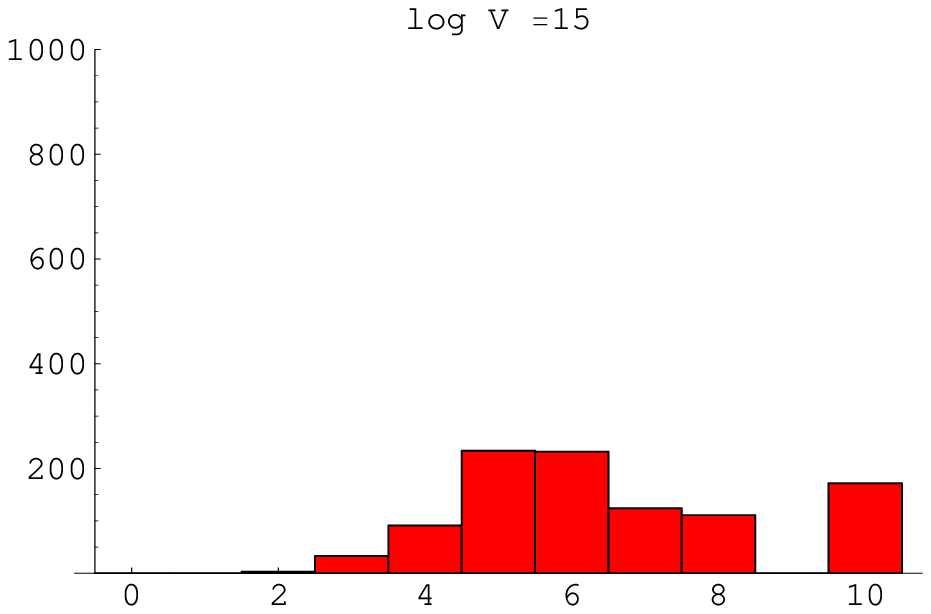} \\
\includegraphics[width=2.5in]{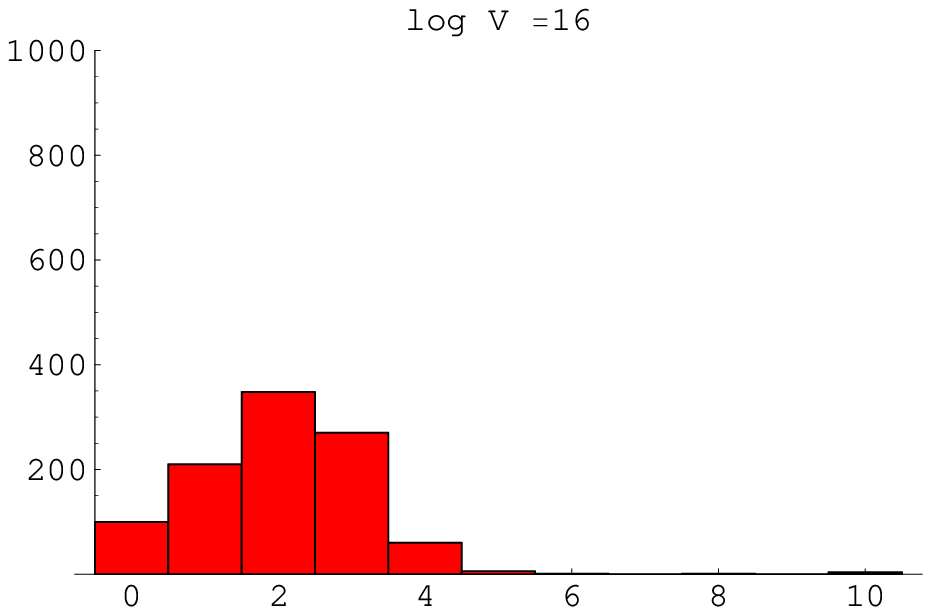} &
\includegraphics[width=2.5in]{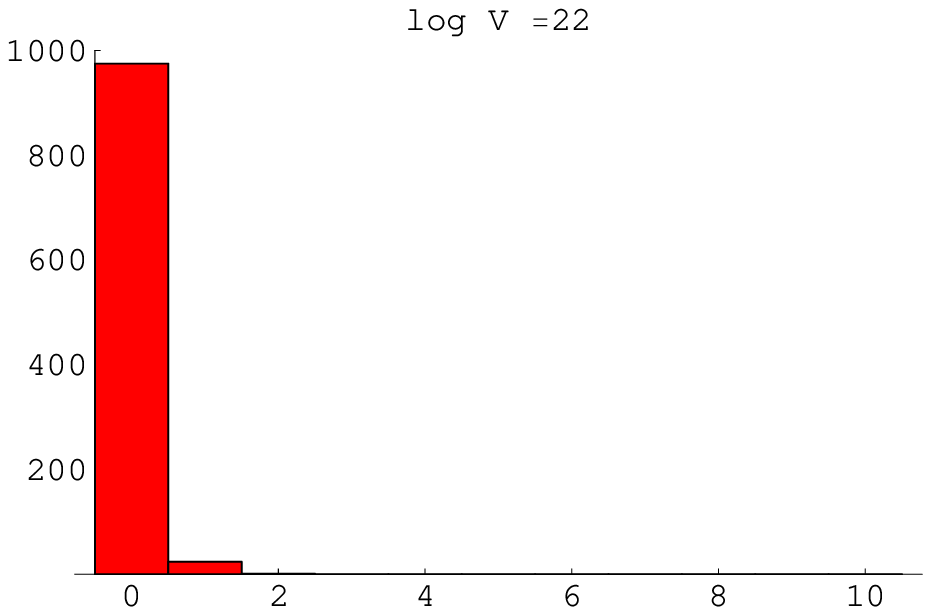}
\end{tabular}
\end{center}
\caption{Probability distribution for the number of unwrapped
dimensions at freeze out, for four different choices of the initial
volume.  The number of unwrapped dimensions is indicated on the
horizontal axis.  Each histogram is a Monte Carlo based on $10^3$
different sets of initial conditions.  Note that it's impossible to
have nine unwrapped dimensions, since the wrapping matrix is
symmetric.}
\end{figure}

As can be seen in Figure 1, larger initial volumes mean fewer
unwrapped dimensions at late times. This can be understood as follows.
For fixed $\dot{\lambda}$ the total area in branes at the start of the
simulation, given by (\ref{Nequi}), is proportional the volume $V$.
If we assume a roughly isotropic universe then the expected number of
branes in each wrapping state $N_{ij} \sim V^{4/5}$, as can be seen in
(\ref{Nijequi}).  Thus larger initial universes will have larger
initial wrapping numbers.  At the start of the simulation the
annihilation rate (\ref{Gammaij}) scales like
\be
\Gamma_{ij} \sim {1 \over V^{3/5}} N_{ij} \sim V^{1/5}\,.
\ee
Thus larger universes are initially more efficient at getting rid of
their branes.  But as the wrapping numbers drop the $V^{-3/5}$
prefactor in the annihilation rate wins out, and larger universes
ultimately find it more difficult to get rid of their branes before
freeze-out.\footnote{The time to freeze-out is roughly independent of
the initial volume and thus does not affect this conclusion.}
Conversely, if the universe starts with a small initial volume the
initial wrapping numbers will be small.  One could easily have all
$N_{ij} < 0.5$, in which case we would regard the initial state as
having no branes present.

\begin{figure}
\begin{center}
\includegraphics[width=4in]{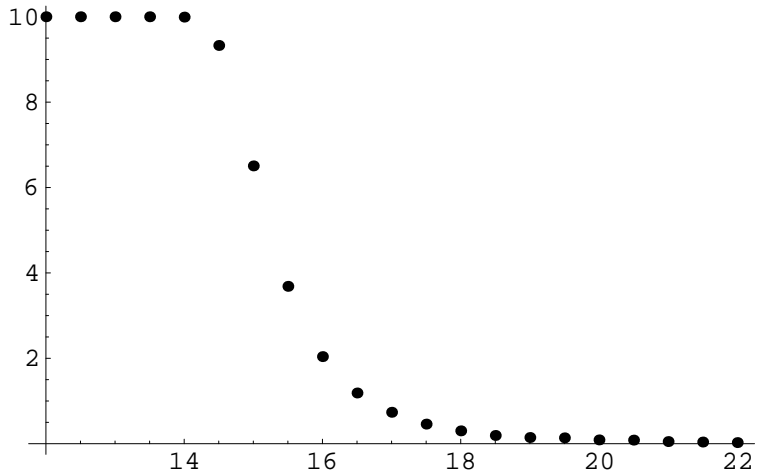}
\end{center}
\caption{Mean number of unwrapped dimensions at freeze-out (y-axis)
versus log of the initial volume (x-axis).}
\end{figure}

Looking at Figure 1, we see that the initial volume determines the
distribution of dimensionality.  For very small volumes, the branes
always annihilate before freeze-out and all ten dimensions unwrap. For
large volumes all directions tend to be wrapped at freeze-out.  Figure
2 shows the transition between these two extremes, by plotting the
mean number of unwrapped directions at freeze-out as a function of the
initial volume.

\subsection{Velocity dependence}

\begin{figure}
\begin{center}
\begin{tabular}{cc}
\includegraphics[width=2.5in]{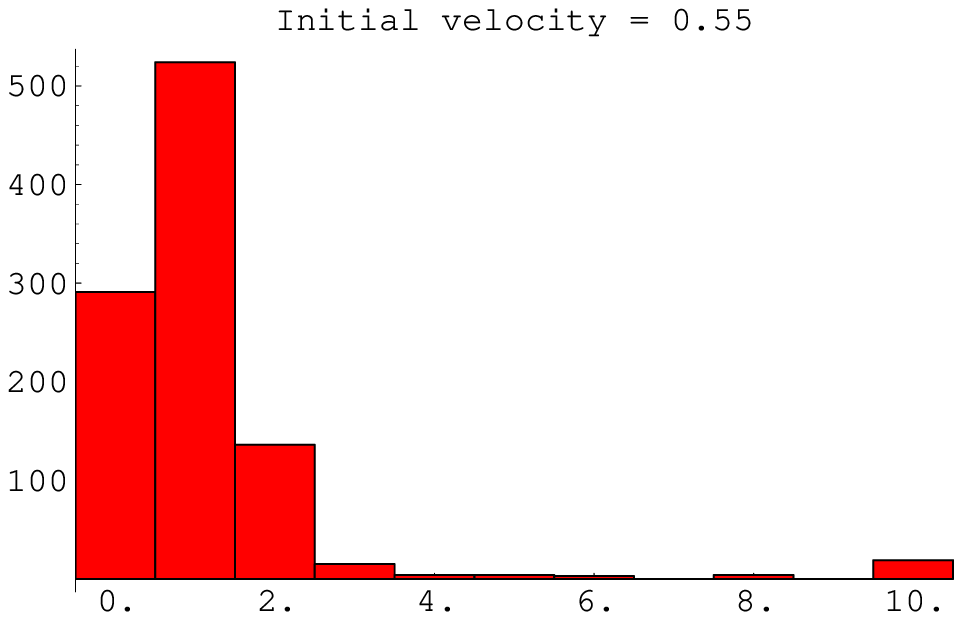} & 
\includegraphics[width=2.5in]{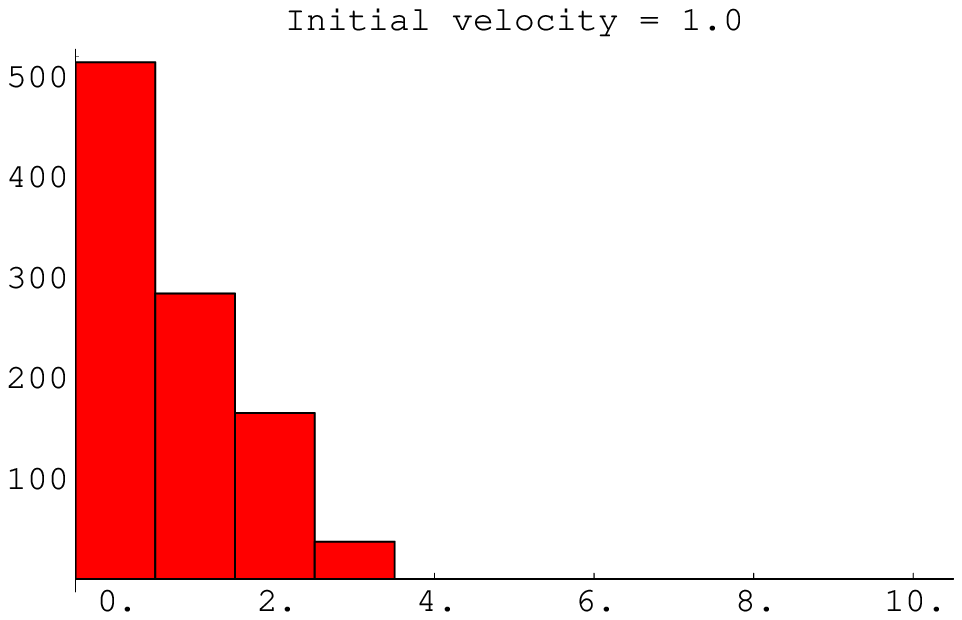} 
\end{tabular}
\end{center}
\caption{Probability distribution for the number of unwrapped
dimensions at freeze out for two different choices of the initial
velocity.  The initial conditions are all $\dot{\lambda}_i = 0.55$
(left plot) and all $\dot{\lambda}_i = 1$ (right plot).  In both plots
the initial volume is fixed to $\log{V} = 20$.  The plots are Monte
Carlos based on $10^3$ different sets of initial conditions.  There is
relatively weak dependence on the initial velocity, as long as
$\dot{\lambda}$ is large enough to start in the Hagedorn phase.}
\end{figure}

\begin{figure}
\begin{center}
\includegraphics[width=4in]{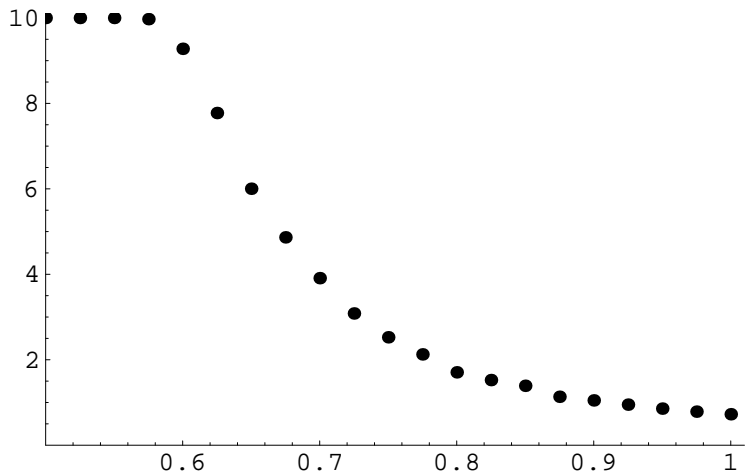}
\end{center}
\caption{Mean number of unwrapped dimensions at freeze-out (y-axis)
versus initial velocity (x-axis).}
\end{figure}

The dependence on initial velocity is much weaker than the dependence
on initial volume, provided the universe is expanding fast enough (and
thus has sufficient energy) to be in the Hagedorn phase when the
simulation begins.  We take all $\dot{\lambda}_i$ to be identical at
the outset, and choose the values of $\lambda_i$ randomly, subject
only to the volume constraint.  Figures 3 and 4 show that the
distribution in the number of unwrapped dimensions at freeze-out only
depends weakly on the initial velocity.  Note that the left hand panel
in Figure 3 shows the distribution for initial velocities which are
only marginally above the value (\ref{MinHagH}) needed to ensure we
start in the Hagedorn phase.  In Figure 5 we show the dependence of
the mean number of unwrapped dimensions on both the initial radii and
velocities.  One can see both the onset of a Hagedorn phase at
$\dot{\lambda} \approx 0.5$, and the volume dependence of the final
number of unwrapped dimensions.

\begin{figure}
\begin{center}
\includegraphics[width=5in]{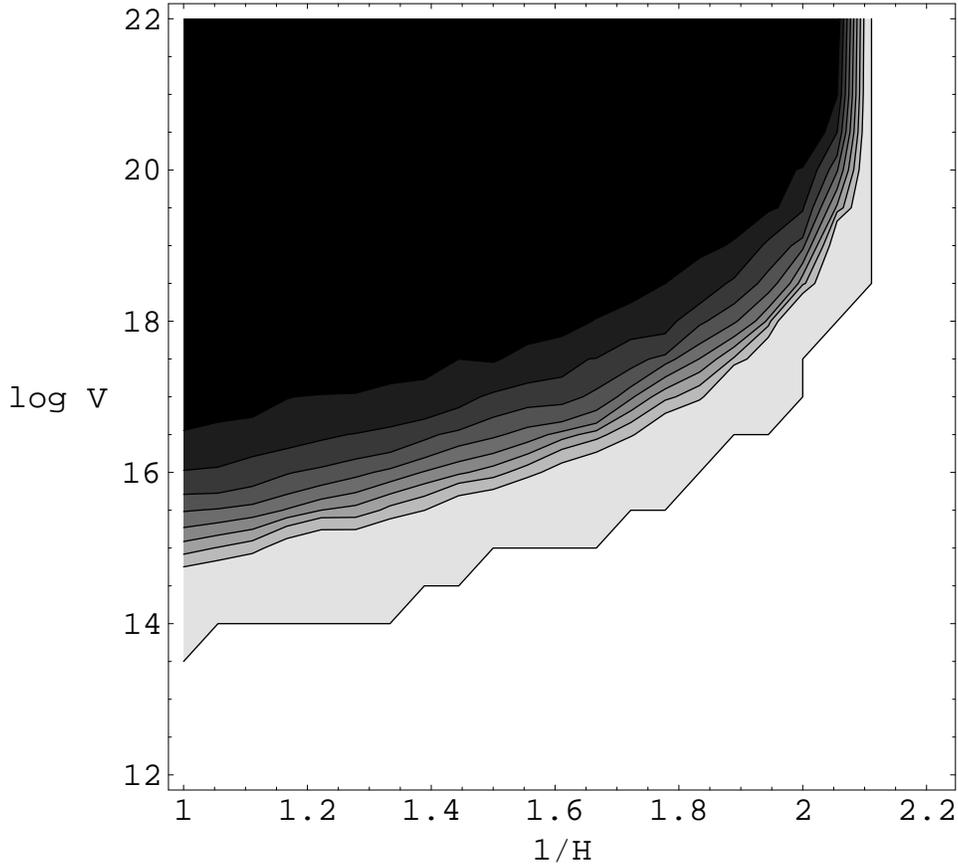}
\end{center}
\caption{This contour plot shows the mean number of unwrapped
dimensions as a function of both the log of the initial volume
(y-axis) and the inverse initial velocity $1/\dot{\lambda}_i \equiv
1/H$ (x-axis).  For each run the initial $\lambda_i$ are chosen
randomly, but the initial $\dot{\lambda}_i$ are all identical.  There
is little dependence on $1/H$, provided we are in the Hagedorn phase
to begin with ($\dot{\lambda} > 0.502$).  The darkest shading
corresponds to a mean number of unwrapped dimensions less than unity,
while the lightest shading corresponds to a mean of 10 (fully
unwrapped).}
\end{figure}

\subsection{Summary}

From these numerical results we see no evidence for a preferred number
of unwrapped dimensions at late times.  Rather the wrapping numbers at
freeze-out depend on the initial conditions.  We therefore cannot
uniquely predict the effective dimensionality of the universe at late
times, although we can assign a probability to different final states.

We can ask what parameters determine the final state.  Provided the
initial volume is large enough, the most important factor determining
the number of unwrapped directions at freeze-out is the anisotropy in
the initial values of the $\lambda_i$. We verified this by running the
code for the same set of $\lambda_i$ and different values of the
$N_{ij}$ (chosen randomly, as outlined above), and found that for
sufficiently large $\log{V}$ the same set of $\lambda_i$ typically
produced the same number of the unwrapped directions, independent of
the values of the $N_{ij}$.
 
Consequently, if we want to sharpen our prediction for the number of
unwrapped dimensions, we must constrain the initial conditions.  A
similar problem arose in the original work of Brandenberger and Vafa
\cite{Brandenberger:1988aj}, where it was argued that a thermal
fluctuation could produce a universe with either one, two, or at most
three large dimensions.  In the Brandenberger-Vafa scenario one could
imagine deploying an anthropic argument to argue against observation
of less than three dimensions.\footnote{For a dynamical approach see
\cite{Watson:2002nx}.}  Rather than pursue anthropic arguments, in the
next section we turn our attention to a rationale for restricting the
space of initial conditions, namely insisting that the initial state
of the universe be consistent with holography.

\section{Holography and initial conditions}

The holographic principle \cite{'tHooft:gx,Susskind:1994vu} is thought
to be a fundamental property of quantum gravity.  Loosely speaking, it
requires that the number of degrees of freedom in a given volume scale
like the surface area.  But so far we have treated M-theory
semi-classically.  In this approximation the number of degrees of
freedom is extensive in the volume, so we run the risk of violating
holography.

We now apply the holographic principle to brane gas cosmology.  For
simplicity we specialize to the case of a square torus ($\lambda_1 =
\cdots = \lambda_{10} = \lambda$) with uniform wrapping ($N_{ij} = N$
for all $i \ne j$).  We will argue that holography is satisfied
provided we put restrictions on the initial conditions.  By combining
holographic bounds with entropy arguments, we will argue for a
preferred set of initial conditions for the universe.  Moreover these
preferred initial conditions have the right qualitative features to
drive the brane gas scenario.

\subsection{Holographic bounds}

The holographic principle was first applied to cosmology in
\cite{Fischler:1998st,Easther:1999gk,Veneziano:1999ts,BakRey}.  We
will use the covariant form of the entropy bound developed in
\cite{Bousso:1999xy,Bousso2} to obtain limits on the initial size of
the universe.  Our analysis closely follows section 3.4 of
\cite{Bousso2}.

In the brane gas scenario one expects that at early times the universe
is in a Hagedorn phase, with scale factor $e^\lambda \sim t^{2/d}$.
At intermediate times there could be a radiation-dominated phase, with scale
factor $\sim t^{2/(d+1)}$.  Finally at late times the universe is
dominated by brane tension; for uniform wrapping this means $e^\lambda
\sim t^{2/(d-2)}$ \cite{Easther:2002qk}.  Thus to a good approximation
throughout its history the universe has a flat FRW metric with a
power-law scale factor.

A flat FRW universe has an apparent horizon at a proper radius $d_{AH}
= 1/\dot{\lambda}$ \cite{BakRey}.  Holography requires that the
entropy inside a spherical volume of radius $R < d_{AH}$ be bounded by
$A/4G$.  That is, for a given entropy density $s$ the radius must
satisfy
\be
s V \leq A / 4 G
\ee
or equivalently
\be
R \leq R_{\rm max} = {d \over 4Gs}
\ee
where we have used the relation $V = RA/d$ appropriate to a sphere in
$d$ dimensions.  If the sphere is larger than the apparent horizon $R
> d_{AH}$ then holography puts no restrictions on the allowed entropy
\cite{Bousso2}.

The entropy density in the Hagedorn phase is given in (\ref{sHag}),
while in the radiation-dominated phase the entropy density comes just
from the supergravity gas:
\be
s = {d+1 \over d} c_S T^{10} = {d+1 \over d} c_S \left({d(d-1) \over
16 \pi G c_S} \dot{\lambda}^2 \right)^{10/11}\,.
\ee
Thus the holographic bound on the radius is
\be
R_{\rm max} = \left\lbrace
\begin{array}{ll}
{d^2 \over 4G(d+1)c_S}\left({d(d-1) \over 16 \pi G c_S}
\dot{\lambda}^2 \right)^{-10/11} & \quad \dot{\lambda} < 0.502 \\
{dT_H \over 4G} \left({1 \over d} c_S T_H^{11} + {d(d-1) \over 16 \pi
G} \dot{\lambda}^2\right)^{-1} & \quad \dot{\lambda} > 0.502
\end{array}
\right.
\ee
For $\dot{\lambda} < 0.234$ it turns out that $R_{\rm max}$ is larger
than the radius of the apparent horizon $d_{AH}$, so holography puts
no restriction on the physical volume of the universe.  For
$\dot{\lambda} > 0.234$, on the other hand, $R_{\rm max}$ is smaller
than $d_{AH}$ and we must limit the size of the universe to satisfy
$e^\lambda \leq 2 R_{\rm max}$.  These bounds are illustrated in
Fig.~6.

\begin{figure}
\epsfig{file=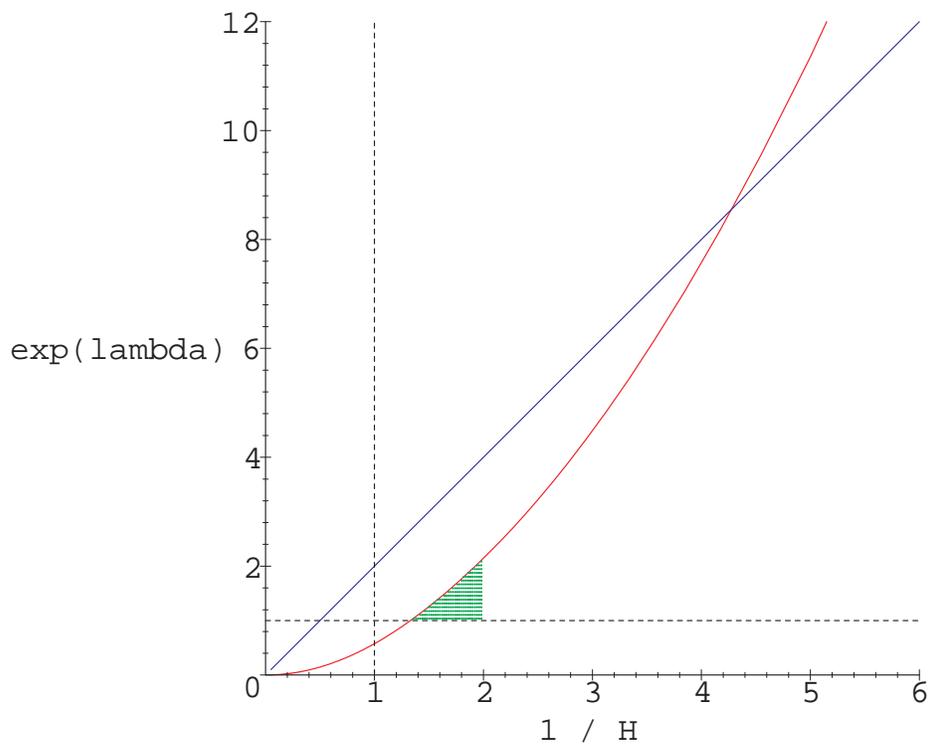}
\caption{Bounds on the size of the universe as a function of $1/H
\equiv 1/\dot{\lambda}$.  The red curve is the holographic bound $2
R_{\rm max}$, the straight blue line is the diameter of the apparent
horizon $2/\dot{\lambda}$, and semiclassical gravity breaks down at
the dotted black lines.  In the green shaded region the universe
satisfies the holographic bound and is in the Hagedorn phase.}
\end{figure}

\subsection{Holography and initial conditions}

We conclude with some speculation about holography and the choice of
initial conditions for the universe.  The basic point is very simple.
Fischler and Susskind have shown that the holographic principle is
satisfied in the universe today \cite{Fischler:1998st}.  Moreover
Flanagan, Marolf and Wald have shown that if holography is satisfied
at some instant of time then it will be satisfied both in the future
and (by time reversal) in the past, up to the point where
semiclassical general relativity breaks down
\cite{FlanaganMarolfWald}.  Evolving our universe backwards in time,
this means the holographic bound must be satisfied until general
relativity breaks down.  This would occur when either the volume is
too small ($e^\lambda \sim 1$) or the Hubble parameter is too large
($\dot{\lambda} \sim 1$).  These bounds are indicated by the dotted
lines in Fig.~6.

Thus the universe must have originated from the horizontal dotted line
in Fig.~6, somewhere to the right of the holographic bound.  Now
consider the expression for the entropy density in the Hagedorn phase
(\ref{sHag}).  The entropy density is an increasing function of
$\dot{\lambda}$, and the total volume is fixed.  Thus on entropic
grounds {\em the preferred initial conditions for the universe
saturate the holographic bound at the point where general relativity
breaks down}.  This behavior was first noted in
\cite{Fischler:1998st}.  Here we are arguing that it is a general
feature.

If this argument is correct, the preferred initial conditions for the
universe have the right qualitative features to drive brane gas
cosmology: the universe begins with a small initial volume and large
initial Hubble parameter (or equivalently a large initial energy
density).  The numerical values suggested by our analysis are
discouraging, unfortunately: the preferred initial conditions are
roughly $\lambda = 0$ and $\dot{\lambda} = 3/4$.  Comparing Figures 5
and 6, we see that the the Hagedorn region identified in Figure~6 is
corresponds to a volume too low to even be plotted in Figure~5.  Given
such small initial volumes very few branes are present in the initial
state.  Thus the most likely evolution of the universe leads to ten
unwrapped and roughly isotropic dimensions.

On the face of it, this is an extremely discouraging result for the
brane gas scenario, as it appears to imply that the initial number of
wrapped branes is very small -- effectively one is starting from a
brane gas without branes.  However there are several reasons for
qualifying this conclusion. First, the comparison of initial volumes
is sensitive to ${\cal O}(1)$ numerical coefficients.  Our use of a
low-energy supergravity action seems to capture the right qualitative
behavior of the Hagedorn phase, but we do not expect it to precisely
capture all numerical coefficients.  For example our estimate of the
M-theory Hagedorn temperature is only an estimate, which surely
receives ${\cal O}(1)$ corrections.  Indeed we expect such corrections
in order to get the M2 and M5 critical temperatures to agree.
Likewise our estimate for the Hagedorn equation of state presumably
receives ${\cal O}(1)$ corrections.

A more fundamental issue is that our discussion of holography assumed
an isotropic torus, while the brane gas scenario relies on an initial
anisotropy to seed the asymmetric growth of dimensions.  It would thus
be interesting to study holographic bounds on an anisotropic torus.
Indeed there is reason to think that in the limit of extreme
anisotropy, where M-theory reduces to IIA string theory, the
holographic bound could be less restrictive, simply because the
holographic bound involves the Planck length while the entropy density
in the stringy Hagedorn phase is set by the string scale.

We should also consider the impact of inhomogeneity on our
analysis. Following Easther and Lowe \cite{Easther:1999gk}, we can
regard the holographic bound as a manifestation of the generalized
second law of thermodynamics. From this perspective, which takes a
more limited view of holography than positing it as a key feature of
some underlying fundamental theory, violations of the holographic
bound are only important if they can be exploited to form a black hole
that contains less entropy than the material used to create it -- thus
violating the generalized second law.  While investigating an
inhomogeneous 11 dimensional spacetime is a forbidding prospect, we
can perform a rough check by asking whether a Schwarzschild black hole
with a mass equal to the entire energy budget of the universe would
fit neatly inside our torus.  For a universe in the Hagedorn phase the
answer is ``no,'' which suggests that homogeneity is not such a bad
assumption.\footnote{We are grateful to Erick Verlinde for a valuable
discussion on this point.}

Even if a more detailed analysis of holography made it possible to
decompactify three dimensions, we would still be faced with a
fine-tuning problem.  That is, our analysis shows that in the M-theory
context the initial volume must fall within a fairly narrow window in
order to have a significant probability of decompactifying three
dimensions.  Let us be optimistic and suppose that by obtaining the
correct numerical coefficients and including the effects of anisotropy
and inhomogeneity we would find that this window overlapped with the
holographically allowed range of initial conditions.  We would still
face the difficulty that small changes in the initial volume
significantly affect the probability of decompactifying three
dimensions.  Part of the appeal of the brane gas scenario was the hope
that a brane gas in the early universe would automatically lead to
decompactification of three dimensions.  In the M-theory context this
hope is not realized.

\section{Conclusions}

In this paper we extended the brane gas scenario in several
directions.  We gave a simple estimate of the Hagedorn temperature for
2-branes, and investigated the properties of a universe dominated by a
Hagedorn gas of 2-branes.  We estimated the cross-section for
interactions between the branes and the SUGRA gas, showing that
annihilation of branes becomes less efficient as the universe expands.
Thus the branes eventually freeze out, leading to a relic density of
winding branes at late times. We numerically solved the corresponding
Boltzmann equations, and found that the number of ``wrapped''
dimensions at late times was essentially determined by the initial
volume, provided the universe starts out in the Hagedorn phase.

We then looked more closely at constraints on the initial conditions,
identifying regions of initial condition space that are compatible
with the holographic bound.  With this cut, we found that in order to
be consistent with holography the initial volume of the universe had
to be relatively small, implying that the equilibrium number of branes
was also small.  Comparing this bound with the numerical work, we saw
that the holographically allowed region of initial condition space
typically leads to a universe in which all branes annihilate before
freeze-out, thereby leaving all ten dimensions free to expand
isotropically.

This result is noteworthy for two reasons. Firstly, it suggests that
the M-theoretic version of the brane-gas scenario cannot produce a
universe with anisotropic distributions of winding branes, and thus
does not provide a mechanism for ensuring that the universe contains a
small number of macroscopic dimensions.  Secondly, it is -- to our
knowledge -- the first time a holographic bound has been applied to a
toroidal cosmology and, more importantly, it is the first time that
holographic arguments have been used to successfully put new
constraints on cosmological models.

These results come with the caveat that while the dynamics were
analysed in a fully anisotropic spacetime, the holographic argument
was formulated for an isotropic universe. It is not clear to us
whether a moderate level of anisotropy could modify the holographic
constraint to the point where decompactification of three dimensions
becomes possible. Moreover, the current analysis involves constants of
order unity that are not reliably determined, providing another
possible loophole in our conclusions.  Finally, we have assumed that
the universe is homogeneous, with branes that are ``smeared out'' in
the transverse directions. This should be a good approximation when
dealing with a large number of branes, but the number of branes we see
are small.

At this point, however, our inclination is to take these results
seriously, and to explore their consequences.  One promising
possibility is to posit that one direction of the torus is small
compared to the Planck scale. This reduces the mass of branes wrapped
around the small direction -- thus increasing their equilibrium number
density -- while holding the overall volume fixed.  In effect, this
scenario describes the stringy limit of M-theory. Our results do not
directly apply to this regime, since our underlying supergravity
action only makes sense if all ten directions are large compared to
the Planck scale.  But the technology we have developed should be
applicable to string gas cosmology, a subject we intend to analyze in
the future.

\section*{Acknowledgements}
BG and DK and supported in part by DOE grant DE-FG02-92ER40699.  MGJ
is supported by a Pfister Fellowship.  ISCAP gratefully acknowledges
the financial support of the Ohrstrom Foundation.  We wish to thank
Raphael Bousso, Robert Brandenberger, Jacques Distler, Nori Iizuka,
Shiraz Minwalla, Lenny Susskind, Henry Tye and Erick Verlinde for
numerous valuable discussions.

\appendix
\section{Equations of state}
Expressions for the energy densities associated with massless
particles in thermal equilibrium can be found in any statistical
mechanics text.  For a relativistic gas in $d$ spatial dimensions
each degree of freedom has an energy density
\begin{eqnarray*}
{\rm BOSON}: \hspace{0.5in} \rho_b &=& \frac{1}{ (2 \pi)^d} S_{d} d! 
\zeta(d+1) T^{d+1} \\
{\rm FERMION}: \hspace{0.5in} \rho_f &=& \frac{1}{ (2 \pi)^d}  S_{d} (1 - 
2^{-d})  d! \zeta(d+1) T^{d+1}
\end{eqnarray*}
There are three instances where this will be used in our work: the 
supergravity gas, the M2-brane worldvolume, and the M5-brane worldvolume.  
In each application, we must remember to sum the energy densities from 
each degree of freedom.
\subsection{Supergravity gas}
For the supergravity gas we have 128 bosonic and 128 fermionic degrees of 
freedom, all massless, in 10 dimensions.  So:
\be \rho _b = \frac{4725 \zeta(11)}{16 \pi^5} T^{11} \hspace{1in} \rho _f = 
\frac{4833675 \zeta(11)}{16384 \pi^5} T^{11} 
\ee
producing
\be \rho _{S} = 128 \rho_b + 128\rho_f \equiv c_S T^{11}, 
\hspace{0.5in} c_S = \frac{9672075\zeta(11)}{128 \pi^5} 
\ee
%
\subsection{$M2$-brane gas}
The M2-brane has transverse fluctuations which we model as a gas composed 
of massless particles residing on the brane.  There are 8 bosonic and 8 fermionic
degrees of freedom.  Thus
\be \rho _b = \frac{\zeta(3)}{\pi}  T^3 \hspace{1in} \rho _f =
\frac{3 \zeta(3)}{4 \pi} T^3 
\ee
producing
\be
\rho_X = 8 \rho_b + 8\rho_f \equiv c_X T^3 , \hspace{0.5in}  c_X = 
\frac{14 \zeta(3)}{\pi}. 
\ee
%
\subsection{M5-brane gas}
M5-branes have 5 transverse coordinates, a 2-form whose field strength
is self-dual, and all their superpartners.  These produce (again) 8
bosonic and 8 fermionic degrees of freedom, but now in five
dimensions.  Thus
\be
\rho_{M5} = 8 \rho_b + 8 \rho_f \equiv c'_X T^6 , \hspace{0.5in} c'_X = 
\frac{\pi^3}{6}. 
\ee
%

\section{Einstein equations}
We begin with the metric ansatz used in \cite{Easther:2002qk},
\be
ds^2  = -dt^2 + \sum_{i=1}^d R_i(t)^2 d\theta_i^2 
\ee
where the angular coordinates $\theta_i$ run from $0$ to $2\pi$. The
Einstein tensor has the following non-zero components
\begin{eqnarray}
{G^t}_t &=& \frac{1}{2} \sum_{k \neq l} \frac{ {\dot R}_k {\dot R}_l}{R_k 
R_l} \\
{G^i}_i &=& \sum_{k \neq i} \frac{ {\ddot R}_k}{R_k} + \frac{1}{2} \sum_{k 
\neq l} \frac{ {\dot R}_k {\dot R}_l}{R_k R_l} - \sum_{k \neq i} \frac{ 
{\dot R}_k {\dot R}_i }{R_k R_i} 
\end{eqnarray}
(no sum on $i$ on the second line).  For future reference,
\be 
\label{trace}
\sum_i {G^i}_i = (d-1) \sum_{k} \frac{ {\ddot R}_k}{R_k} + 
\frac{d-2}{2} \sum_{k \neq l} \frac{ {\dot R}_k {\dot R}_l}{R_k R_l} 
\ee
The Einstein equations are
\be
G^t{}_t = 8 \pi G \rho \qquad G^i{}_i = - 8 \pi G p_i
\ee
where the energy density and pressures are given in (\ref{rhofull}),
(\ref{pfull}).  At this point it is convenient to set $\theta_i = 2
\pi x_i$ and $e^{\lambda_i(t)} = 2\pi R_i(t)$.  That is, we write the
metric as
\be ds^2 = - dt^2 + \sum_{i=1}^d e^{2 \lambda_i(t)} dx_i^2 \qquad
0 \leq x_i \leq 1. 
\ee
In terms of these variables the Einstein equations are
\bea
\label{einspace}
&& {1 \over 2} \sum_{i \not= j} \dot{\lambda}_i \dot{\lambda}_j = 8 \pi G \rho \\
&& {\ddot \lambda_i} + \frac{{\dot V}}{V} {\dot \lambda}_i =
8 \pi G \left({1 \over d-1} \rho + p_i - {1 \over d-1} \sum_k p_k\right)
\eea
To obtain the second line it is useful to take the sum of the
space-space equations and use (\ref{trace}).

\section{Hagedorn temperatures}

There is a distressing factor of $2^{1/3} \approx 1.26$ between our
limiting M2-brane temperature and that found by Russo
\cite{Russo:2001vh}.  While such a factor would not affect the
qualitative features of our analysis, it is important to determine the
reason for the discrepancy.

In studying the membrane energy, we used the following large-winding
approximation to the membrane mass:
\[
m \sim T_2 A + \hbox{\rm (ideal gas of transverse fluctuations)}\,.
\]
This expression is valid at low temperature (and thus at low
excitation number), but there is no reason to believe it is valid near
the Hagedorn temperature.  Russo identifies the critical temperature
at which a membrane wound on the Euclidean time dimension becomes
tachyonic.  But unlike the present work Russo does not expand the
membrane action for large winding.  We believe this is the origin of
the discrepancy.\footnote{Both Russo and the present work ignore
interactions on the membrane worldvolume.}  The complications of
studying membranes prevent us from showing this directly, but an
analogous numerical discrepancy can be seen in the following string
calculation.

The (exact) bosonic string spectrum is ($\alpha'=1$)
\[
m^2 = w^2 R^2+ N-4
\]
where we let $N$ refer to all oscillations.  If we consider only
unwound strings, this means $m \sim \sqrt{N}$ at high oscillation
number, and the partition function behaves roughly like
\begin{eqnarray*}
Z &=& \sum _N d(N) e^{-E/T} \\
&\sim& \int _0 ^\infty dN \ e^{\sqrt{N} (\beta_H - \beta)} \\
\end{eqnarray*}
which diverges for $T \geq T_H$.  Now consider winding strings, for
which a mass approximation {\it valid at low excitation number} would
be
\begin{equation}
\label{windapprox}
m \sim wR+ \frac{N}{2 wR}\,.
\end{equation}
The partition function is now roughly
\begin{eqnarray*}
Z &=& \sum_{N,w \neq 0} d(N) e^{-E/T} \\
&\sim& \sum _{w \neq 0} \int _0 ^\infty dN \ \exp \left[ {\sqrt{N} \beta_H
- (wR + N/2wR)\beta} \right] \\
&\sim& \sum _{w \neq 0} e^{wR( \beta _H^2/2 \beta - \beta)}
\end{eqnarray*}
which diverges for $T \geq \sqrt{2} T_H$.  This calculation is
incorrect because we used the approximation (\ref{windapprox}) for
arbitrarily large $N$; at any fixed $w$ there will eventually be an
excitation number $N \gg w^2R^2$ at which it is better to use
\[
m \sim \sqrt{N} + \frac{wR}{2\sqrt{N}}
\]
which will reproduce the old divergence at $T=T_H$.  Since each $w$
term in the partition function diverges towards positive infinity at
this temperature, the total partition function will also diverge.
Thus the $2^{1/2}$ discrepancy is merely an artifact of the large
winding approximation to the mass.


\end{document}